\documentclass[fleqn,usenatbib]{mnras}
\usepackage[T1]{fontenc}
\usepackage{ae,aecompl}

\usepackage{graphicx}	
\usepackage{amsmath}	
\usepackage{amssymb}	
\usepackage{braket}
\usepackage{courier}

\newcommand{\erf}{erf}

\title[eigenfunctions of phase space spirals]{Eigenfunctions of Galactic Phase Space Spirals from Dynamic Mode Decomposition}

\author[Keir Darling et al.]{
Keir Darling\thanks{E-mail: 13kd39@queensu.ca}
and Lawrence M. Widrow\thanks{E-mail: widrow@queensu.ca}
\\
Department of Physics, Engineering Physics \& Astronomy, Queen's University, Stirling Hall, Kingston, ON K7L 3N6, Canada
}

\date{Accepted XXX. Received YYY; in original form ZZZ}

\pubyear{2018}

\begin{document}

\label{firstpage}
\pagerange{\pageref{firstpage}--\pageref{lastpage}}
\maketitle

\begin{abstract}

We investigate the spatiotemporal structure of simulations of the
homogeneous slab and isothermal plane models for the vertical motion
in the Galactic disc. We use Dynamic Mode Decomposition (DMD) to
compute eigenfunctions of the simulated distribution functions for
both models, referred to as DMD modes. In the case of the homogeneous
slab, we compare the DMD modes to the analytic normal modes of the
system to evaluate the feasibility of DMD in collisionless self
gravitating systems. This is followed by the isothermal plane model,
where we focus on the effect of self gravity on phase mixing. We
compute DMD modes of the system for varying relative dominance of
mutual interaction and external potential, so as to study the
corresponding variance in mode structure and lifetime. We find that
there is a regime of relative dominance, at approximately $ 4:1 $
external potential to mutual interaction where the DMD modes are
spirals in the $ (z,v_z) $ plane, and are nearly un-damped. This leads
to the proposition that a system undergoing phase mixing in the
presence of weak to moderate self gravity can have persisting spiral
structure in the form of such modes. We then conclude with the
conjecture that such a mechanism may be at work in the phase space
spirals observed in Gaia Data Release 2, and that studying more
complex simulations with DMD may aid in understanding both the timing
and form of the perturbation that lead to the observed spirals.

\end{abstract}



\section{Introduction}

Astrometric and radial velocity surveys of the Milky Way such as SEGUE \citep{segue}, RAVE \citep{rave}, LAMOST \cite{lamost}, and Gaia Data Release 2 (GDR2) \citep{brown2018, gaiadr2}, have revealed a panoply of phase space structures in the Galactic stellar disc.  In the vicinity of the Sun, these structures include vertical asymmetries in the local stellar number density \citep{widrow2012, yanny2013, bennett2018}, vertical bulk motions of disc stars \citep{widrow2012, williams2013,carlin2013,quillen2018,gaiadr2} and phase spirals in $z-v_z$ projections of the stellar distribution function (DF) \citep{antoja2018}. In addition, there is evidence for corrugations of the disc \citep{xu2015} at distances of $10$ to $15$ kpc from the Galactic centre. At even larger radii, there is the prominent warp \citep{binney1992,sellwood2013}, as well as evidence for disc stars kicked up to large Galactic latitudes \citep{pricewhelan2015}. Taken together, these observations point to a disk in a state of disequilibrium.

Perhaps the most intriguing of the aforementioned structures are the phase spirals in the radial velocity subsample of GDR2 \citep{antoja2018}. They were first seen by selecting stars in an arc of about 8 degrees in Galactic azimuth and $200\,{\rm pc}$ in Galactocentric radius centered on the Sun and plotting the number density, mean $v_\phi$ or mean $v_R$ across the $z-v_z$ plane. The spirals have now been studied as a function of position within the disc, action variables, and stellar properties \citep{blandhawthorn2019, laporte2019, shen2019}. 

A heuristic explanation of the spirals is that a local bend in the
disc phase mixes due to the anharmonic nature of the vertical
potential \citep{antoja2018}, while coupling of the vertical and
in-plane motions then leads to the $v_R$ and $v_\phi$ spirals
\citep{binney2018, darling2018}. In the simplest implementation of
this picture, one treats stars as test particles in a fixed
potential. This model seems to capture the basic features of the
spirals and allows one to estimate the time at which the initial
perturbation that gave rise to them took place \citep{antoja2018}.
Nevertheless, it leaves several questions unanswered, which include the
following: What perturbed the disc? Can we point to a singular event
that drove the disc from equilibrium or is the disc in a perpetual
state of disequilibrium? What is the underlying DF of the perturbed
disc. The phase spirals are likely the de-projection of the number
count asymmetry along the $z$ axis mentioned above to the $z-v_z$
plane. Can we understand the spirals as the projection of some
higher dimensional structure?

A number of candidates have been proposed as the agent of
disequilibrium. For example, the disc may have been perturbed by a
passing satellite galaxy or dark matter subhalo with the Sagittarius
dwarf a prime suspect \citep{laporte2019, blandhawthorn2019}. On the
other hand, the buckling of the stellar bar has been shown to generate
phase spirals in simulations of a Milky Way-like galaxy
\citep{khoperskov2019}.

The more challenging problem is to discern the phase space
DF. \citet{tremaine1999} stressed the idea that the dimensionality of
the DF can change via phase mixing. For example, a
satellite galaxy that is being tidally disrupted by the gravitational
potential of its host galaxy changes from a six-dimensional structure
to a three-dimensional stream. \citet{tremaine1999} showed that one
can relate changes in the phase space structure of a system to the
eigenvalues of the Hessian matrix for the Hamiltonian. In principle,
the method could be applied to phase mixing in a Galactic disc.

 The main drawback of the phase mixing arguments is that they ignore the self-gravity of the perturbation, which is clearly important for the development of both bending and density waves in the disc. In short, when a local region of the disc is displaced from the midplane, it exerts a perturbing force on the unperturbed disc that, at least in linear theory, is the same order as the restoring force pulling it back into the midplane due to the unperturbed disc \citep{hunter1969}. A striking example of the importance of self-gravity can be found in the toy model simulations of bending waves in \cite{darling2018} (Figure 7). Of course, self-gravity is built into the simulations of \citet{laporte2019, jbh2019} and \citet{khoperskov2019}.

These considerations suggest that phase mixing and self-gravity are competing effects. A particularly simple toy model in which this competition plays out is the one-dimensional slab. This system can be thought of as an idealized disc in which all of the structure is in the direction normal to the midplane. It can be studied using linear perturbation theory as well as 1D N-body simulations \citep{mathur1990, weinberg1991, widrow2015}. For an isolated system there exists a trivial zero frequency normal mode corresponding to the displacement of the system relative to the midplane as well as a continuous spectrum of modes. If an external restoring force is included, then the displacement mode no longer has zero frequency. Moreover, if the DF for the system is truncated in energy, then gaps open up in the continuum and one has the possibility of additional discrete modes. In general, perturbations of the system will involve a mixture of discrete modes and modes from the continuum where the former should lead to eternal oscillations while the latter phase mix on time-scales related to the gradient of the frequency across the continuum. In his N-body simulations, \citet{weinberg1991} found that it was difficult to excite the discrete modes, presumably because power was leaking into nearby parts of the continuum where phase mixing was occurring. He did find that the system exhibited oscillations that were long-lived as compared to its dynamical time. 

A similar situation arises when one applies linear perturbation theory to a simple two-dimensional model for a galactic disc comprising concentric, rotating, razor-thin rings. This model, which was employed to study warps, is in some sense complementary to the slab model. If the system is isolated or embedded in a spherical halo, there is a zero-frequency mode corresponding to the tilting of the system as a whole. As with the slab model, there is also a continuous spectrum of modes \citep{hunter1969, sparke1988}. On the other hand, if the system is embedded in a flattened halo, then the tilt mode becomes non-trivial and, in fact, has features similar to those seen in warped galaxies \citep{sparke1988}. That said, the discrete warp mode found by \citet{sparke1988} does not appear to exist when the ring-model disc is immersed in a live halo \citep{binney1998}.

Evidently, a proper treatment of stellar dynamics in galactic discs must capture the physics of both phase mixing and self-gravity. In this paper, we propose that dynamic mode decomposition (DMD) can provide a route to achieving this end. DMD was developed in the field of computational fluid dynamics to study problems involving turbulent flows and jets \citep{schmid2010}, and is closely related to Koopman theory (see \cite{mezic2005} and \cite{rowley2009}) It is essentially a dimensionality reduction algorithm for time-series data with the aim of identifying the dominant eigenfunctions of a system. At first glance, eigenvalue methods, which generally require a linear operator, would seem to be incompatible with the nonlinear problem at hand. The idea is to ``lift'' the dynamics from the state space, where the dynamics is governed by nonlinear physics to a space where the dynamics is described by a linear (generally infinite dimensional) operator, usually referred to as the Koopman operator \citep{mezic2005}. DMD can provide a finite dimensional approximation of this operator comprised of dominant eigenfunctions of the system. One is left with a low-dimensional space in which the evolution of the system may be represented linearly and the dominant structure readily studied. DMD is similar in aim to the methods in \cite{tremaine1999} in terms studying structure and its dimensionality. However since DMD is data driven, the inclusion of self gravity is much simpler as compared to the process of obtaining an appropriate action space Hamiltonian.

The layout of the paper is as follows. In Section \ref{oscillations}
we present a summary of the DMD method and its connection with the
theory of small oscillations. In Section \ref{SLAB} we apply DMD to an
N-body simulation of the homogeneous slab model. This system exhibits
small oscillations about an equilibrium state, which can be compared
to analytically derived normal modes
\citep{antonov1971,kalnajs1973}.  We then turn, in Section
\ref{Simulation}, to the isothermal plane, which serves as a simple
model for the vertical structure of a Galactic disc.  Our isothermal
plane simulations are constructed so that we can adjust the relative
importance of self-gravity and an external potential. In this case DMD
is used to study the interplay between phase mixing and oscillatory
modes. In doing so it helps elucidate the physics of $z-v_z$ phase spirals and the
importance of self-gravity.  In Section \ref{extension} we suggest a
path forward for using DMD with full six-dimensional simulation
data. Finally, we conclude with a summary and discussion in Section
\ref{Conclusion}.

\section{Characteristic Oscillations}\label{oscillations}

\subsection{Small Oscillations}\label{smalloscillations}


Consider a classical system with $n$ degrees of freedom described by the generalized phase space coordinates $ \mathbf{x}=(\mathbf{q},\mathbf{p})^T$. In the neighborhood where oscillations of the system are small, we consider a linearized Hamiltonian $H(\mathbf{q},\mathbf{p}) = U(\mathbf{q}) + K(\mathbf{p})$, where both the potential $U(\mathbf{q})$ and the kinetic energy $K(\mathbf{p})$ are quadratic forms \citep{arnold1989}:

\begin{equation}\label{UT}
U(\mathbf{q}) = \tfrac{1}{2}\mathbf{q}^T\mathbf{Bq}, \ \  K(\mathbf{p}) = \tfrac{1}{2}\mathbf{p}^T\mathbf{C}\mathbf{p}.
\end{equation}

\noindent In this case, the equations of motion can be written as a linear operator equation,

\begin{equation}\label{linearsystem}
\frac{d\mathbf{x}}{dt} = \boldsymbol{\mathcal{A}}\mathbf{x},
\end{equation}

\noindent where 

\begin{equation}\label{Ahamiltonian}
\boldsymbol{\mathcal{A}} = \begin{pmatrix}
\mathbf{0} & \mathbf{C} \\
-\mathbf{B} & \mathbf{0} \\
\end{pmatrix}.
\end{equation}

\noindent The solution is then given in terms of the eigendecomposition of $\boldsymbol{\mathcal{A}}$,

\begin{equation}\label{eigenfunctionSolution}
\mathbf{x}(t) = \sum_{j}b_j\boldsymbol{\phi}_je^{\omega_j t},
\end{equation}

\noindent where $ \boldsymbol{\phi}_j $ and $ \omega_j $ are the
eigenvectors and eigenvalues of $ \boldsymbol{\mathcal{A}} $, that is,
$\boldsymbol{\mathcal{A}}\phi_j = \omega_j\phi_j$ and the coefficients
$b_j$ are determined from the initial conditions.

\subsection{Dynamic Mode Decomposition}\label{DMD}

In this section, we provide a compact overview of DMD, which draws
from the introductory chapters of \cite{kutz2016}. We consider a
nonlinear dynamical system that is described by some general state
vector $\mathbf{x}(t)$, which does not necessarily belong to a
Hamiltonian system. The goal of DMD is to determine the best-fit
linear model for the non-linear dynamics. The key idea is that over a
sufficiently short time interval $\Delta t$, the dynamics of the
system can be approximately described by a linear system of equations
of the form given in equation \ref{linearsystem} with a solution given
by equation \ref{eigenfunctionSolution}.

With these considerations in mind, we draw $m$ discrete time samples $ \mathbf{x}_j $ from the system with a sampling period of $ \Delta t $. We then construct the equivalent discretized system of equations

\begin{equation}\label{dtsys}
\mathbf{x}_{j+1} \approx \mathbf{A}\mathbf{x}_j,
\end{equation}

\noindent where the discrete-time map is given by $ \mathbf{A} =
e^{\mathcal{A}\Delta t} $. We emphasize here that the states $
\mathbf{x}_j $ need not be coordinates of the system, but can be any
set of observables.

In general, the operator $ \mathbf{A} $ is not known but is approximated from the data. To do so, we construct the data matrix

\begin{equation}\label{X}
\mathbf{X} = \begin{pmatrix}
| & | & & | \\
\mathbf{x}_1 & \mathbf{x}_2 & ... & \mathbf{x}_{m-1} \\
| & | & & | \\
\end{pmatrix},
\end{equation}

\noindent and the time shifted data matrix

\begin{equation}\label{X'}
\mathbf{X'} = \begin{pmatrix}
| & | & & | \\
\mathbf{x}_2 & \mathbf{x}_3 & ... & \mathbf{x}_{m} \\
| & | & & | \\
\end{pmatrix}.
\end{equation}

\noindent Our system of equations can then be approximated with the matrix equation

\begin{equation}\label{linear}
\mathbf{X'} \approx \mathbf{A}\mathbf{X}.
\end{equation}

\noindent From this, $ \mathbf{A} $ is estimated by minimizing the matrix norm, $||\mathbf{X'}-\mathbf{A}\mathbf{X} ||$, which yields the result 

\begin{equation}\label{A}
\mathbf{A}  = \mathbf{X'}\mathbf{X}^+.
\end{equation}

\noindent Here, $\mathbf{X}^+$ denotes the Moore-Penrose pseudo-inverse of $ \mathbf{X} $, which can be computed via singular value decomposition (SVD). As in \cite{numrec} the SVD of $ \mathbf{X} $ is defined by the relation $\mathbf{X} \approx \mathbf{U}\mathbf{\Sigma}\mathbf{V}^\dagger $ where $ ^\dagger $ denotes the Hermitian conjugate transpose, $ \mathbf{\Sigma} $ holds the singular values along its diagonal, and $ \mathbf{U} $ and $ \mathbf{V} $ are comprised of left and right orthonormal vectors respectively. Using the SVD, we have that $ \mathbf{X}^+ = \mathbf{V}\mathbf{\Sigma}^{-1}\mathbf{U}^\dagger $, with which the discrete-time map becomes

\begin{equation}\label{Asvd}
\mathbf{A} \approx \mathbf{X'}\mathbf{V}\mathbf{\Sigma}^{-1}\mathbf{U}^\dagger.
\end{equation}


\noindent Our next goal is to obtain the eigendecomposition of $ \mathbf{A} $ as we did with $ \mathbf{\mathcal{A}} $ in Section \ref{smalloscillations} to facilitate understanding the time evolution of the system in terms of dominant modes. In the spirit of principal component analysis, we assume that the dominant structure of the system may be described by $ r < m $ modes. Recall that in principle component analysis, when a data matrix $ \mathbf{X} $ possesses low dimensional structure, it may be reasonably approximated in a basis spanned by the $ r $ column vectors in $ \mathbf{U} $ of its SVD corresponding to the $ r $ largest singular values. We therefore work with the projection of $ \mathbf{A} $ into this $r$-dimensional subspace, 

\begin{equation}\label{Atilde}
\tilde{\mathbf{A}} = \mathbf{U}^\dagger\mathbf{A}\mathbf{U}=\mathbf{U}^\dagger\mathbf{X'}\mathbf{V}\mathbf{\Sigma}^{-1}. 
\end{equation}

\noindent By working with this projection, we drastically reduce the dimension of the discrete-time map, making its eigendecomposition computationally tractable despite the typically large data matrices. Doing so also improves the numerical stability of the pseudo-inverse of $\mathbf{X}$. 

We now determine the eigenvalues and eigenvectors of $\tilde{\mathbf{A}}$. That is, we solve the equation $\tilde{\mathbf{A}}\mathbf{\Xi}=\mathbf{\Xi}\mathbf{\Lambda}$ where $\mathbf{\Lambda}$ is a diagonal matrix whose elements $\lambda_j$, $j = 1,\dots r$ are the eigenvalues of $\tilde{\mathbf{A}}$, and $\mathbf{\Xi}$ is a matrix whose columns are the corresponding eigenvectors. To a good approximation, the $r$ most dominant eigenvalues of $\mathbf{A}$ are the $\lambda_j$ while the corresponding eigenvectors, which are often referred to as the DMD modes, are given by 

\begin{equation}\label{Phi}
\mathbf{\Phi} = \mathbf{X'}\mathbf{V}\mathbf{\Sigma}^{-1}\mathbf{\Xi} = \begin{pmatrix}
| & | & & | \\
\boldsymbol\phi_1 & \boldsymbol\phi_2 & ... & \boldsymbol\phi_{r} \\
| & | & & | \\
\end{pmatrix}.
\end{equation}

\noindent (For a detailed explanation and proof, see \citet{tu2013}.)

We now have all of ingredients necessary to write a series solution for the state of the system. The solution takes the form of equation \ref{eigenfunctionSolution}, where as before $b_j$ are the initial amplitudes of the modes, given by $ \mathbf{b} = \mathbf{\Phi}^+\mathbf{x}_1 = \big(b_1 \ ... \ b_r \big)^T $,  and the frequencies are $ \omega_j = \ln(\lambda_j)/\Delta t $.

In general, $\lambda_j$ can be real, imaginary, or complex. The case
$\lambda_j=1$ ($\omega_j = 0$) corresponds to a time-independent mode
and arises, for example, when one has a system that is oscillating
about some equilibrium configuration. Imaginary $\lambda_j$ indicates
a pure oscillating mode, what would usually be referred to as a normal
or true mode of the system. Real $\lambda_j$ correspond to pure
growing or decaying modes while complex $\lambda_j$ correspond to pure
damped or growing oscillations. It is often convenient to split the
modal decomposition into terms with real and complex eigenvalues. In
particular, for a system described by some real function, the DMD
eigenvalues come in complex conjugate pairs and the associated pairs
of modes combine to yield real contributions in the modal
decomposition.

We conclude this Section with a few remarks that relate DMD back to our earlier discussion of small oscillations in Section \ref{smalloscillations}.
When DMD is applied to a system that is solvable by the method of characteristic oscillations, say, a linearized system with $n$ degrees of freedom, it is natural to construct the data matrices from measurements of the $n$ generalized coordinates. One can then use the full discrete-time map $\mathbf{A}$ and expect to obtain the $n$ characteristic or normal modes of the system. The power of DMD becomes manifest when we consider complex, nonlinear systems where simple analytic methods fail. In such cases, we can construct the data matrices using some convenient set of observables, where the $r$ modes that one obtains are data rather than model driven.

\section{The Homogeneous Slab}\label{SLAB}

In this section, we apply DMD to the linear oscillations of a
self-gravitating, collisionless system of particles about its
equilibrium state. For pedagogical reasons, we take the equilibrium
state to have uniform density within a prescribed distance from the
origin. The system was first studied by \citet{antonov1971} and
\citet{kalnajs1973}. Its simplicity derives from the fact that in the
equilibrium state, all particles undergo simple harmonic motion about
the origin. Furthermore, the linear modes are discrete and purely
oscillatory. In addition, the DF for these modes can be expressed as
elementary functions of the phase space variables, which we take to be
$z$ and $v_z$. By contrast, the oscillations of the isothermal plane
considered in Section \ref{Simulation}  include discrete and continuum modes, which
can be purely oscillatory or damped and can only be derived
numerically using complex analysis.

\subsection{Analytic Considerations}

By Jeans theorem, the DF for an equilibrium, 
one-dimensional system can be written as a function of the sole
integral of motion, the energy $E = v_z^2/2 + \psi(z)$.  For the
homogeneous slab, the DF is given by

\begin{equation}\label{SLABf}
f_0^K(E) = \begin{cases}
2^{-3/2}\pi^{-2}\big(1-E\big)^{-\frac{1}{2}}, \quad 0<E<1 \\
0, \quad \text{otherwise},
\end{cases}
\end{equation}

\noindent where the superscript $ K $ signifies that this is the original Kalnajs model. We use dimensionless units such that the
velocity dispersion, extent of the system in $z$, and Newton's
constant are all set to unity \citep{widrow2015}. In these units, the
system has uniform density $ \rho_0=(2\pi)^{-1} $ in the region
$|z|\leq1 $, and zero density for $ |z|>1 $. All particles
undergo simple harmonic motion about $ z=0 $ at a frequency of $
\Omega_c=\sqrt{2} $. Thus the transformation to action angle variables
is analytic and has a simple geometric interpretation:

\begin{equation}\label{AAtransformation}
\begin{aligned}
z=E^{\frac{1}{2}}\cos(\theta), \ \ v_z = (2E)^{\frac{1}{2}}\sin(\theta)~.
\end{aligned}
\end{equation}

\noindent Note that the phase space distribution is bounded by the
ellipse, $ E=z^2 + v_z^2/2 = 1 $.

The normal modes of the system can be derived directly from a set of
density-potential pairs \citep{antonov1971, kalnajs1973,widrow2015},
which removes the usual need for the Kalnajs matrix method \cite{}.
Letting $ P_n $ be the Legendre
polynomial of degree $ n $, the density and potential of $ j\text{th}
$ mode are written

\begin{equation}\label{slabPotential}
\psi_j^K(z) = N_j\big(P_{j+1}(z)-P_{j-1}(z)\big),
\end{equation}

\noindent and 

\begin{equation}\label{rhoj}
\rho_j^K(z) = -\tfrac{1}{4\pi}\,\tfrac{j(j+1)}{1-z^2}\,N_j\big(P_{j+1}(z)-P_{j-1}(z)\big),
\end{equation}

\noindent where $N_j = \big(\frac{2\pi}{2j+1}\big)^\frac{1}{2}$. 

We focus here on the lowest order even and odd parity modes, which
correspond to $ j=1 $ and $ j=2 $, respectively. From equation
\ref{slabPotential} we have

\begin{equation}\label{key}
\begin{aligned}
\psi_1^K(z) &= \frac{3N_1}{2}\big(z^2-1\big) \\
\psi_2^K(z) &= \frac{5N_2}{2}\big(z^3-z\big).
\end{aligned}
\end{equation}

\noindent With use of the appropriate Legendre polynomials and
trigonometric identities, these are written in terms of the
action-angle variables as

\begin{equation}\label{slabPotential2}
\begin{aligned}
\psi_1^K(z) &= \frac{3N_1}{2}\bigg(E\cos^2(\theta)-1\bigg) \\
\psi_2^K(z) &= \frac{5N_2}{2}\bigg(\frac{E^\frac{3}{2}}{4}\cos(3\theta)+\bigg(\frac{3E^\frac{3}{2}}{4}-E^\frac{1}{2}\bigg)\cos(\theta)\bigg).
\end{aligned}
\end{equation}

\noindent Since both potentials are even in the periodic
variable $\theta$, we may write them as even Fourier series,
distinguishing the even and odd parity modes by their Fourier
coefficients. That is,

\begin{equation}\label{fourierSeries}
\begin{aligned}
\psi_1^K(z) &= \sum_{n>0}\psi_{1,n}\cos(n\theta), \ \ n \in 2\mathbb{Z}+1 \\
\psi_2^K(z) &= \sum_{n>0}\psi_{1,n}\cos(n\theta), \ \ n \in 2\mathbb{Z}.
\end{aligned}
\end{equation}

\noindent Comparison of equations \ref{slabPotential2} and
\ref{fourierSeries} implies that the relevant Fourier coefficients are

\begin{equation}\label{fourierCoefficients}
\begin{aligned}
\psi_{1,2} &= \frac{3N_1E}{4}, \\ 
\psi_{2,1} &= \frac{5N_2}{2}\bigg(\frac{3E^\frac{3}{2}}{4}-E^\frac{1}{2}\bigg), \ \ \psi_{2,3} = \frac{5N_2}{2}\bigg(\frac{E^\frac{3}{2}}{4}\bigg).
\end{aligned}
\end{equation}

\noindent As in \cite{widrow2015} we can also write the mode DFs as a Fourier series

\begin{equation}\label{fj}
Re\{f_j\} = \frac{df_0}{dE}\sum_n \frac{2\psi_{j,n}\cos(n\theta)}{2-\omega^2/n^2},
\end{equation}

\noindent where the sum over $ j $ is restricted to even values for
the even parity modes, and odd values for the odd parity
modes. We therefore have

\begin{equation}\label{f1f2}
\begin{aligned}
Re\{f_1^K(E,\theta)\} &= \frac{df_0^K}{dE}\frac{\psi_{1,2}\cos(2\theta)}{1-\omega^2/8} \\
Re\{f_2^K(E,\theta)\} &= \frac{df_0^K}{dE}\bigg(\frac{2\psi_{2,1}\cos(n\theta)}{2-\omega^2} +\frac{2\psi_{2,3}\cos(3\theta)}{2-\omega^2/9}\bigg),
\end{aligned}
\end{equation}

\noindent where the derivative term $ df_0^K/dE $ is restricted to the
domain $ 0\leq E<1 $, as it is infinite at the boundary $ E=1
$. Finally, we obtain $\rho_j$ by integrating equation \ref{f1f2} over
$v_z$. Consistency with equation \ref{rhoj} then leads to a $j$'th order
polynomial
equation $\omega^2$, which can be solved to obtain
$\omega/\Omega_c=1.73 $ for $ f_1 $, and $ \omega/\Omega_c=1.07, 2.80
$ for $ f_2 $ \citep{kalnajs1973}. Note that in general, there are $j$ modes
for a given $j$ and we therefore have a double series of  modes.

\subsection{DMD Modes of the Homogeneous Slab}\label{slabSimulation}

We next simulate the homogeneous slab model. The initial conditions are 
a sample of $N=2\times 10^5$ equal mass particles drawn from the equilibrium
DF. The system is first evolved for 80 dynamical times. As we will see, during
this period it settles into a new equilibrium state in which the 
discontinuities in the DF and density are smoothed out. Note that the 
only source of perturbations is the shot noise from the finite number of
particles. The system is then evolved for an additional 80 dynamical times
and it is this period of the simulation that we use for our DMD analysis.

Snapshots comprise the DF $f(z,\,v_z)$ estimated on a $150\times 150$
grid in the $z-v_z$ plane for $\{|z|, |v_z|\} < 2$ and are sampled at
a frequency of $10\Omega_c$. To estimate the DF, we take the number of
particles in each phase space cell, multiply by the particle mass, and
divide by the phase space volume of the cell.  Each snapshot is then
reshaped into a single column vector. The set of snapshots are then
combined to yield the data matrices $\mathbf{X}$ and
$\mathbf{X}'$. These matrices have shapes $22,500\times 800$ and
therefore satisfy one of the conditions of DMD, namely that they be
``tall and skinny'' \citet{kutz2016}. The DMD solution is then
computed with a rank of $ r=35 $.

\begin{figure}
	\centering  
	\includegraphics[width=9.2cm]{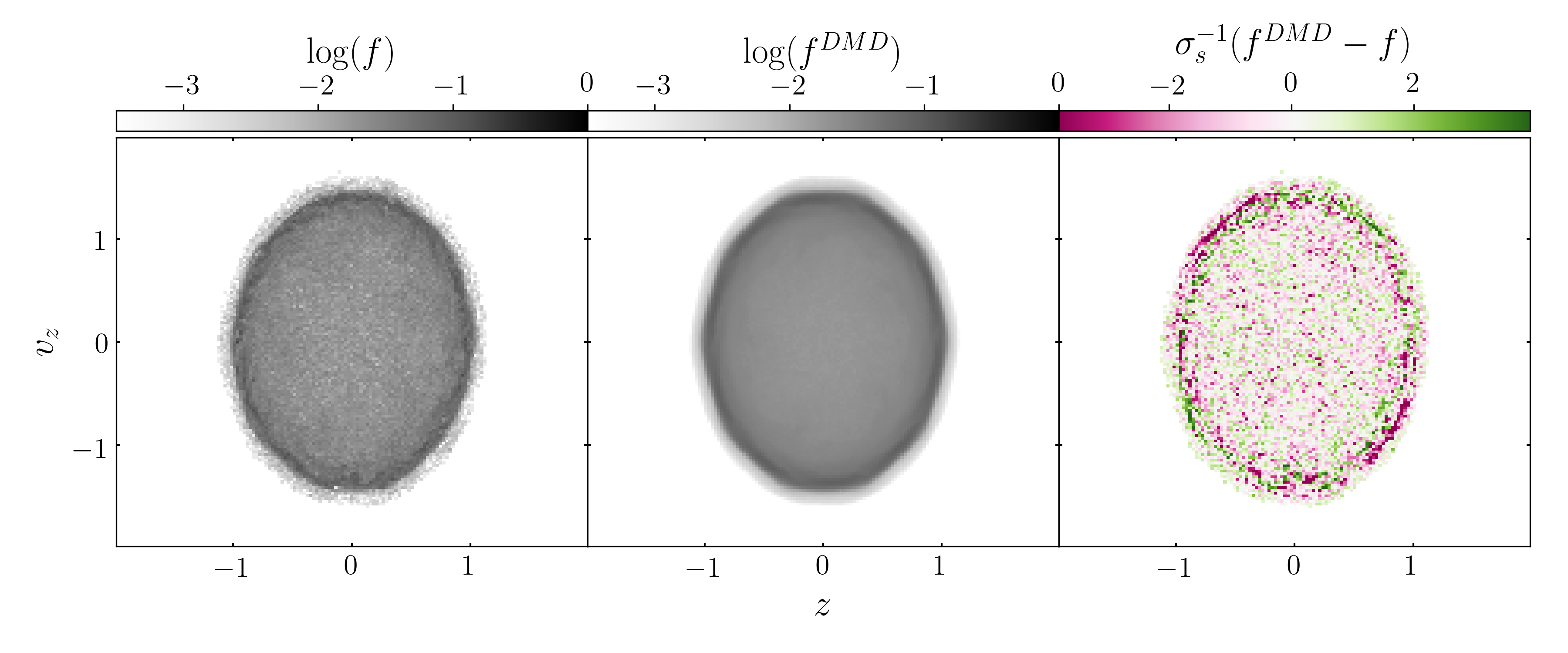}
	\caption{A single snapshot of the slab simulation, the
		computed DMD solution, and the corresponding residual. We
		show only a single snapshot here, as the snapshots of the
		system do not vary in a visually observable way in the
		studied time interval. The error is divided by the magnitude
		of shot noise fluctuations, $ \sigma_s $ to highlight the
		error in the DMD solution independent of noise in the
		simulation.}\label{timeevolutionslab}
\end{figure}

Fig.\,\ref{timeevolutionslab} shows a single snapshot from halfway through the simulation period used for the DMD analysis. In the first panel, we show an
estimate for the DF.  As noted above and discussed in more detail
below, the discontinuity at $E=1$ is smoothed out. In the middle panel
we show the DMD solution while in the final panel, we show the
difference between the two in units of the uncertainty, estimated from
root-N statistics. The residual structure is dominantly random and the
errors in each bin lie within the expected range of fluctuations.
There are larger residuals for $E\sim 1$, which we believe correspond
to high order structure in the simulation that is not accurately
captured by our relatively low-rank DMD solution. Since the residuals lie within the statistical expectation, we take the DMD model to be accurate enough to be proceed.

Following Section \ref{DMD}, and noting that we have a real valued
observable, we split the DMD series solution into separate sums over
real and complex modes:

\begin{equation}\label{fseries}
f^{DMD}(z,v_z,t) = \sum_{j=1}^{r-q}b_j\boldsymbol{\phi}_j e^{\omega_j
	t} + \sum_{k = -\frac{q}{2}}^{\frac{q}{2}}b_k\boldsymbol{\phi}_k
e^{\omega_k t},
\end{equation}

\noindent where the the $ q $ complex eigenfunctions and eigenvalues
satisfy $ \boldsymbol{\phi}_k^* = \boldsymbol{\phi}_{-k}$ and $
\lambda_k^* = \lambda_{-k} $ respectively. The linear combination of
two modes that belong to a conjugate pair and are weighted by complex
amplitudes $ a_k(t)=b_ke^{\omega_kt} $ is real.  That is, for a
conjugate pair of DMD modes, we define the corresponding mode of the
DF to be

\begin{equation}\label{key}
f_j^{DMD}(z,v_z,t) = \sum_{l=\pm k} b_{l}\boldsymbol{\phi}_{l}e^{\omega_{l}t}.
\end{equation}

\begin{figure}
	\centering  
	\includegraphics[width=8cm]{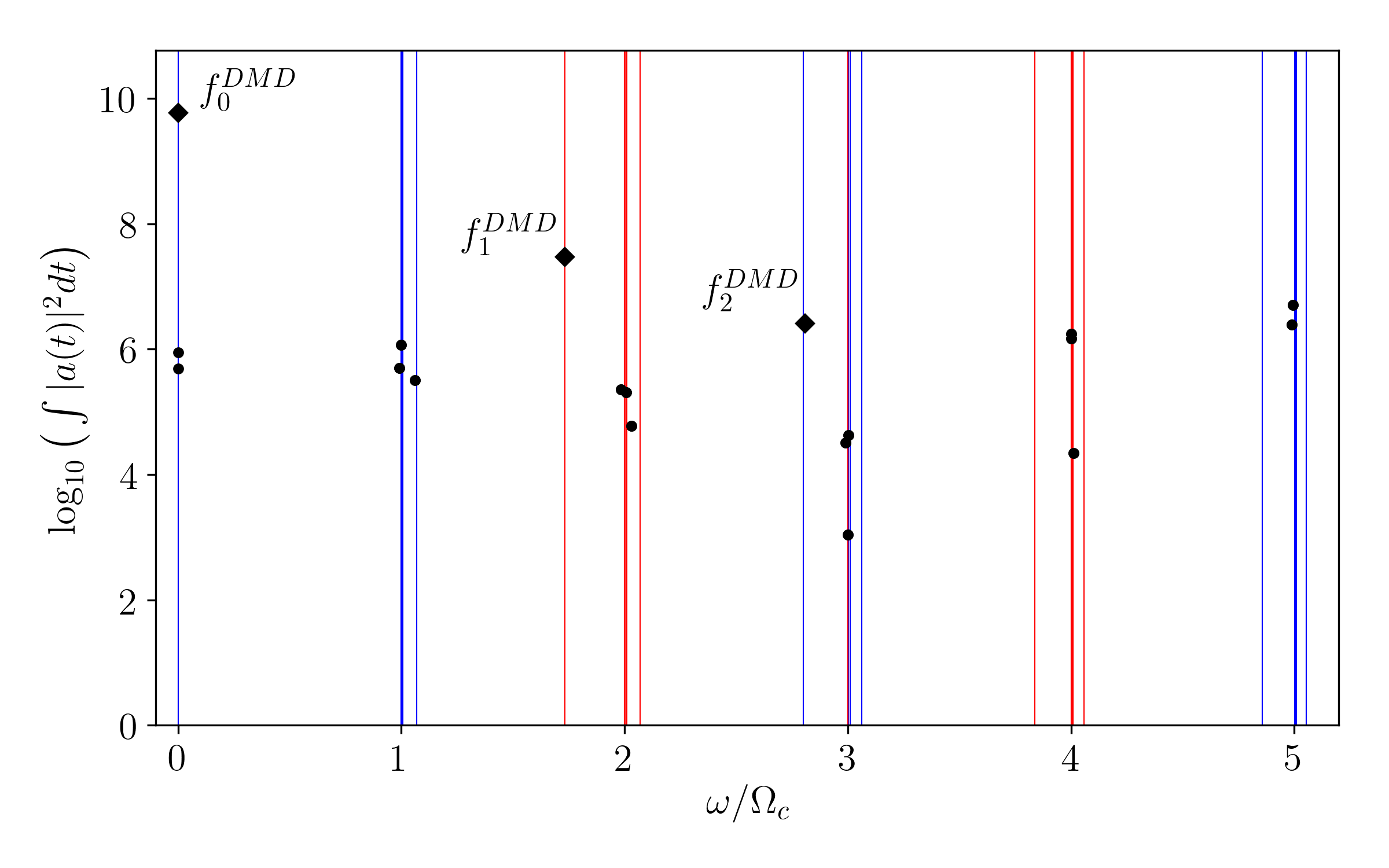}
	\caption{Mode amplitude signal energy as a function of mode
		frequency. The black points correspond to DMD modes, and the
		lines are plotted at the theoretical frequencies of the
		Kalanjs slab model, with blue and red lines indicating even
		and odd parity respectively.}\label{spectrum}
\end{figure}

In Fig. \ref{spectrum} we show the signal energy of
the mode amplitudes as a function of mode frequency, alongside lines
indicating the theoretical frequencies of the slab model from
\cite{kalnajs1973}. Reassuringly, agreement between the DMD mode
frequencies and the expected ones is excellent.

\begin{figure}
	\centering  
	\includegraphics[width=8cm]{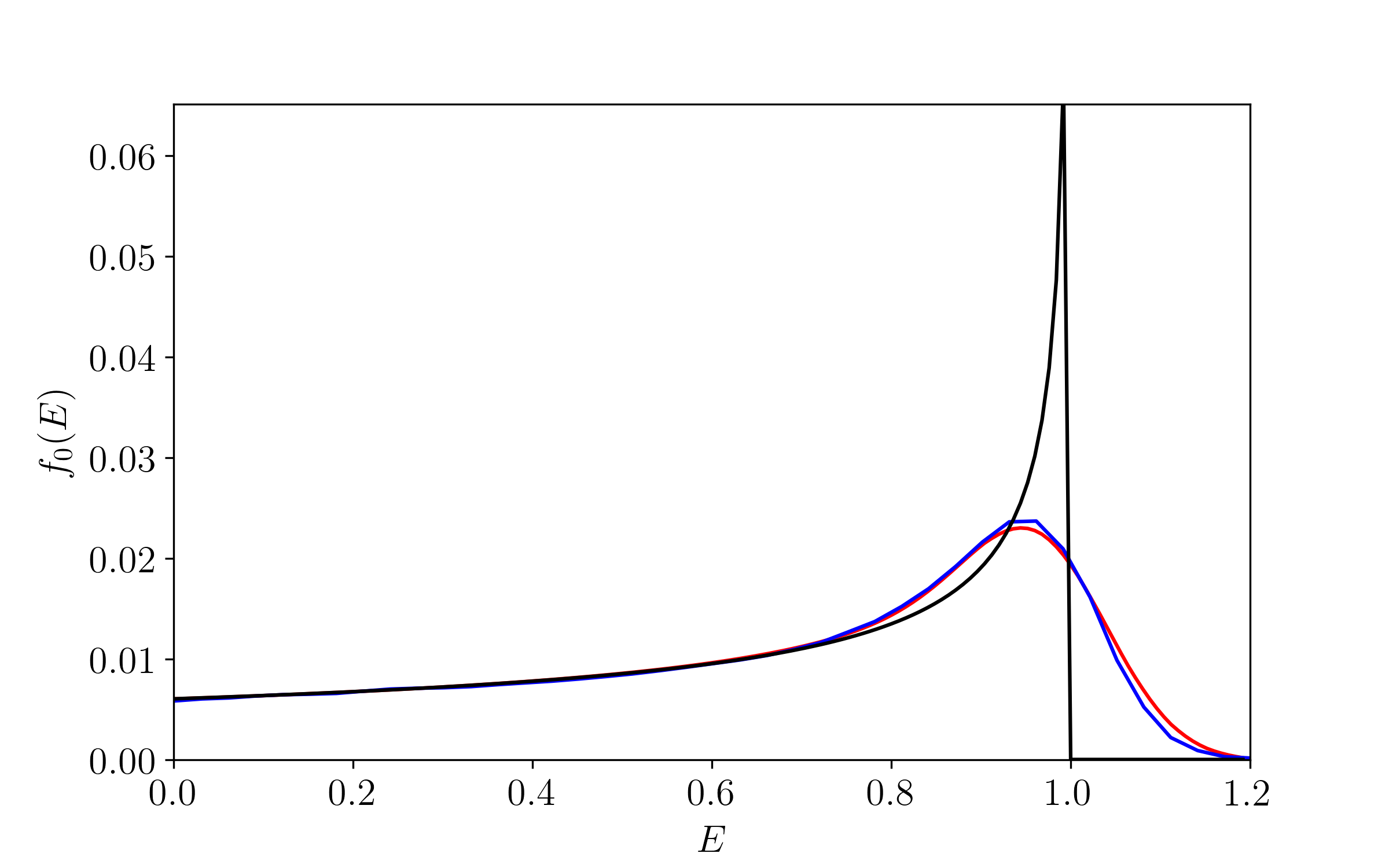}
	\caption{Equilibrium DFs of the Kalnajs model (black), modified slab model (red), and the equilibrium DMD mode (blue). In computing $ f_0^{DMD}(E) $ we have assumed the potential corresponding to the fitted density in equation \ref{modifiedDensity}.}\label{SLABdfs}
\end{figure}

We now focus on a few particular DMD modes beginning with the highest
amplitude mode. This mode has zero frequency, as seen in
Fig.\,\ref{spectrum} and corresponds to the equilibrium state. Its
energy distribution $f_0^{DMD}(E)$, found by integrating the DF over
the angle variable $\theta$, is shown in Fig.\,\ref{SLABdfs}. As
mentioned above, the discontinuity in the energy distribution near
$E=1$ is smoothed out.  (Note that in constructing $f_0^{DMD}(E)$ we
use the potential for the original homogeneous slab model, which turns
out to be an excellent approximation to the potential for the smoothed
distribution.)

\begin{figure}
	\centering  
	\includegraphics[width=9cm]{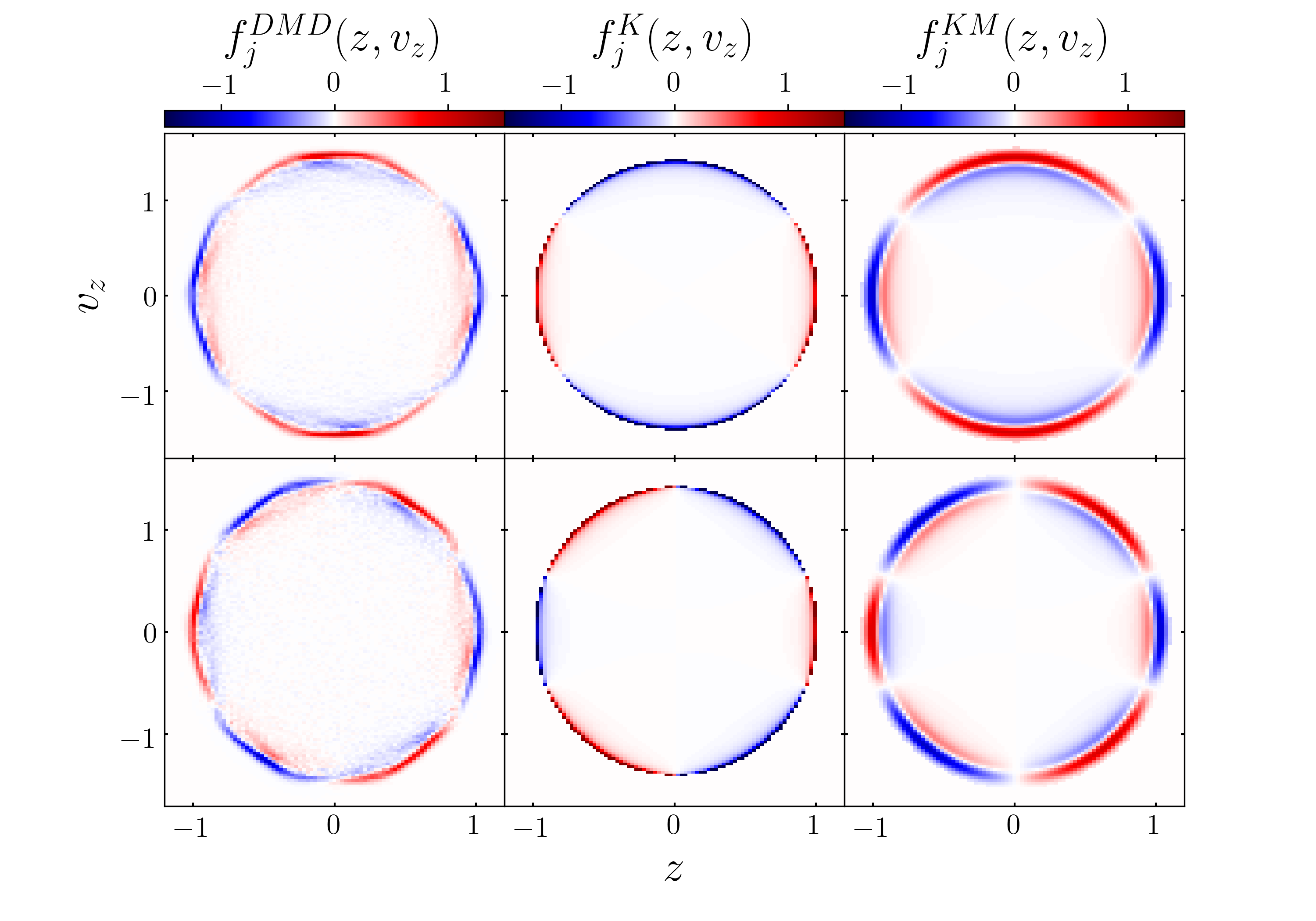}
	\caption{Mode DFs of the DMD solution, the Kalnajs model, and the modified slab model. The normalization of the three models has been adjusted, and a rotation has been applied to the DMD modes for easy comparison.}\label{slabModeComparison}
\end{figure}

We next turn to the $\omega/\Omega_c=1.73,~j=1$ and
$\omega/\Omega_c=2.80,~j=2$ modes. The analytic DFs for these modes
are given in equation \ref{f1f2} and plotted in the middle panels of
Fig.\,\ref{slabModeComparison}. The frequencies and amplitudes of the
corresponding DMD modes are shown as large diamonds in
Fig.\,\ref{spectrum} while the DMD modes themselves are plotted on the
lefthand panels of Fig.\,\ref{slabModeComparison}. The main difference
between the analytic DFs and the DMD modes is found just beyond $E=1$.
Evidently, the DMD modes extend beyond $E=1$ while the analytic modes
have a sharp edge there.

\subsubsection{Modified Slab Model}

The discrepancies between DMD modes and analytic linear modes are
clearly related to the $df_0/dE$ factor in equation \ref{fj}. To
explore the discrepancies further we return to our discussion of the
zero frequency mode. As seen in Fig.\,\ref{SLABdfs}, the sharp peak
and discontinuity in $f(E)$ is smoothed out in the
simulation. Likewise, the discontinuity in the density is also
smoothed out. In fact, the density is well represented by the fitting
formula

\begin{equation}\label{modifiedDensity}
\rho_0^{KM}(z) = \frac{1}{4\pi}\bigg(1+\erf{\bigg(\frac{1-|z|}{a}\bigg)}\bigg),
\end{equation}

\noindent where the constant $ a $ corresponds to the width of the
transition region from $\rho_0=1/2\pi$ to $0$. For our simulation, we
find a best-fit value of $ a=0.037 $. A smoothed out homogeneous slab model was presented in \citet{binney1993}, although with a different functional form that what we use here. The potential associated with
equation \ref{modifiedDensity} can be derived from the Greens function
for the Poisson equation in one dimension and can be expressed in
terms of special functions.

Armed with this density-potential pair, we can then derive the associated 
equilibrium DF via the Abel transform

\begin{equation}\label{AbelIntegral}
f(E)=
-\frac{\sqrt{2}}{\pi}\int_{E}^{\infty}\frac{d\rho}{d\psi}\frac{d\psi}{\big(\psi-E\big)^\frac{1}{2}}.
\end{equation}

\noindent The integrand in equation \ref{AbelIntegral} approaches
infinity when the energy and potential take on similar values. To
remedy this, we make a change of variable letting $ \gamma =
\sqrt{2}(\psi-E)^{\frac{1}{2}} $, such that equation \ref{AbelIntegral}
becomes

\begin{equation}\label{key}
f(E) = -\frac{2}{\pi}\int_{0}^{\infty}\frac{d\rho}{d\psi}d\gamma.
\end{equation}

\noindent This integral must be evaluated numerically.
The resulting DF is shown in Fig. \ref{SLABdfs}, together with $f(E)$ for
the homogeneous slab (equation \ref{SLABf}) and the zeroth order DMD
mode. As can be seen in this figure, the equilibrium distribution
function of the modified slab has a finite and smooth derivative that
changes sign in the neighborhood of $ E=1 $, but otherwise behaves
similarly to that of the homogeneous slab model.

The righthand panels of Fig. \ref{slabModeComparison} show the linear
modes (equation \ref{fj}) with $df_0/dE$ calculated from our smoothed
DF. They are qualitatively similar to the DMD modes in that they
change sign for $E\simeq 1$ and extend beyond the $ E=1 $ ellipse.

In summary, for this simulation, DMD has constructed a set of modes
that include a zeroth order equilibrium distribution and linear
oscillatory perturbations in a manner reminiscent of perturbation
theory.  However, unlike perturbation theory, the zeroth order and
linear perturbations are calculated simultaneously and directly from
simulation data, without direct appeal to the underlying physics
(i.e., the linearized Boltzmann and Poisson equations). In addition,
the DMD analysis does not require that the perturbations be
small. Indeed, the analysis is perfectly applicable to simulations of
nonlinear systems.

\section{The Isothermal Plane}\label{Simulation}

We now consider the isothermal plane model, first developed by \cite{spitzer1942} and \cite{camm1950}, and used as an approximation for the vertical structure of a stellar disc by \cite{freeman1978} and \cite{vanderkruit1981}. 

\subsection{Model Details}

In this model, the equilibrium DF and density are given by

\begin{equation}
f_{\rm eq}\left (z,\,v_z\right ) = \frac{\rho_0}{\left (2\pi\sigma_z^2\right )^{1/2}} 
e^{-E_z/\sigma_z^2}
\end{equation}

\noindent and

\begin{equation}
\rho_{\rm eq}(z) = \rho_0 e^{-\psi(z)/\sigma_z^2},
\end{equation}

\noindent where $ E_z $ is the vertical energy, and $ \sigma_z $ is the velocity dispersion. For an isolated, self-gravitating system, the density and potential must satisfy the Poisson equation and we have 

\begin{equation}
\psi_{\rm eq}(z) = 2\sigma_z^2 \ln{\cosh{\left (z/z_0\right )}},
\end{equation}

\noindent where $\rho_0 = \sigma_z^2/2\pi Gz_0^2$. 

Here we split the gravitational force into two parts: a time-independent part, coming from masses external to the disc, such as the dark halo, and a live part coming from the disc itself. That is, we write the potential as

\begin{equation}
\psi(z,t) = \psi_{\rm ext}(z) + \psi_{\rm live}(z,t),
\end{equation}

\noindent where $\psi_{\rm ext} = \left (1-\alpha\right )\psi_{\rm eq}$ and $\psi_{\rm live}$ comes from the disc with masses reduced by a factor of $\alpha$ relative to what they would be in the isolated case. Thus $ \alpha $, which we call the live fraction, quantifies the relative dominance of self gravity and the external potential. In equilibrium, the total potential is just $\psi_{\rm eq}$, but once the system is perturbed, $\psi_{\rm live}$, $\rho$, and $f$ all depend on time. For definiteness, we use $\sigma_z = 20\,{\rm km\,s}^{-1}$ and $z_0 = 500\,{\rm pc}$, which yields a surface density of $\Sigma = 2z_0\rho_0 = 60\,M_\odot\,{\rm pc}^{-2} $. 

\begin{figure}
	\centering  
	\includegraphics[width=9.2cm]{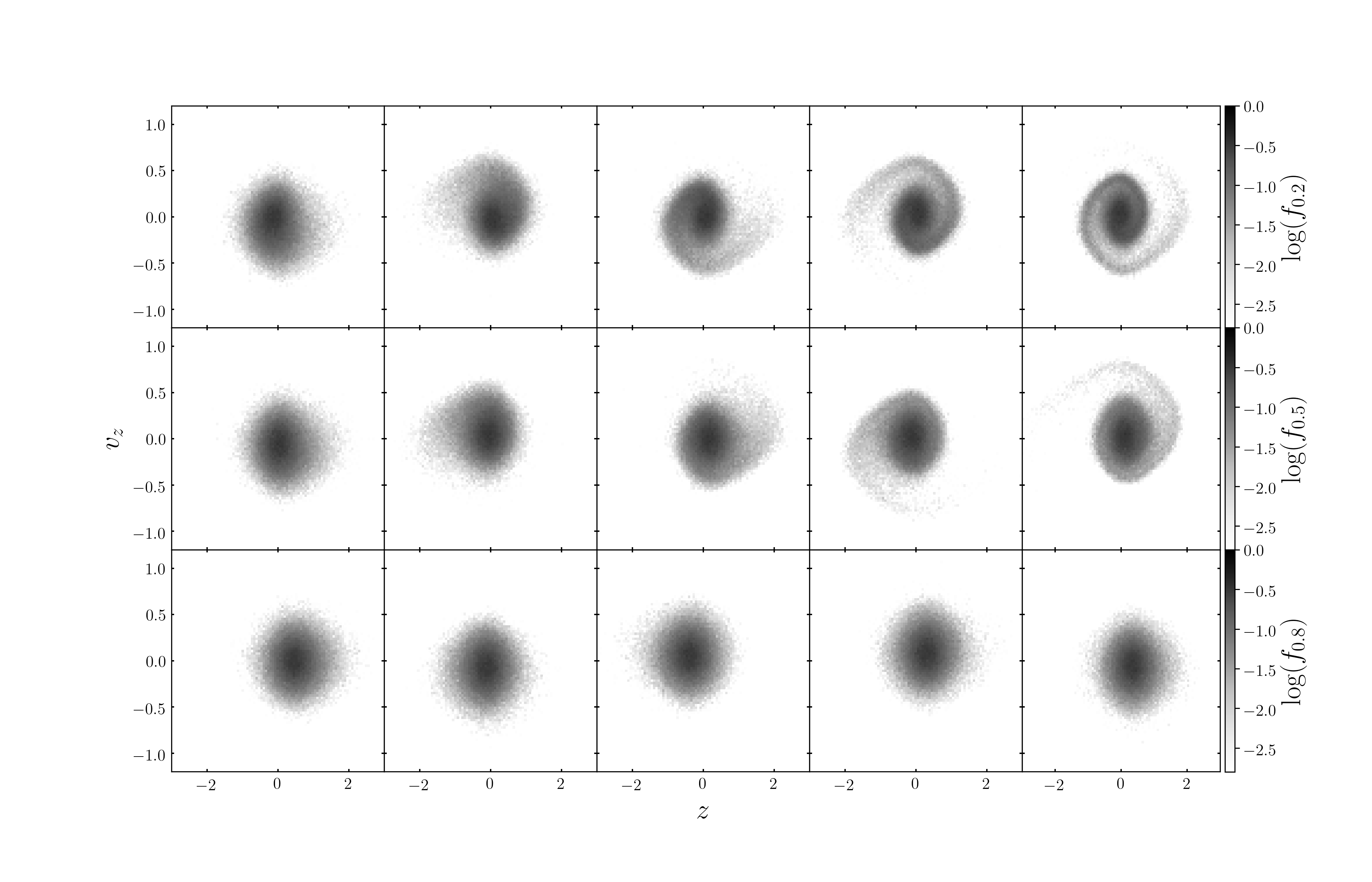}
	\caption{Snapshots of $ \log_{10}(f(z,v_z))  $ for $ \alpha = 0.2, 0.5, 0.8 $ from top to bottom. Time increases in intervals of $ 85  \text{Myr} $ from left to right. The $ z $ axis is in units of kpc, and the $ v_z $ axis is in units of $ 100 \ \text{kms}^{-1}$. The bin sizes are $ \Delta z = 0.079$ kpc and $ \Delta v_z = 0.032 \ \text{km}\text{s}^{-1}$. The same units and bin sizes are used in all following $ (z,v_z) $ projections.}\label{timeevolution}
\end{figure}

To simulate this system we sample $ N=10^5 $ particles from $ f_{\rm eq} $ and then impose a simple bending wave perturbation by shifting the velocities $ 10 \ \text{km}\,\text{s}^{-1} $. This form of perturbation has been shown to yield spirals in the $ (z,v_z) $ phase space similar to that observed in Gaia DR2 \citep{antoja2018, darling2018, binney2018}.  We then evolve the distribution for four orbital periods, or approximately $ 450 \ \text{Myr} $. The time evolution of the phase space density for the cases $\alpha=0.2,\,0.5$ and $0.8$ are shown in Fig. \ref{timeevolution}. 

\subsection{DMD Modes of the Isothermal Plane}\label{dfmodes}

\begin{figure}
	\centering  
	\includegraphics[width=9.5cm]{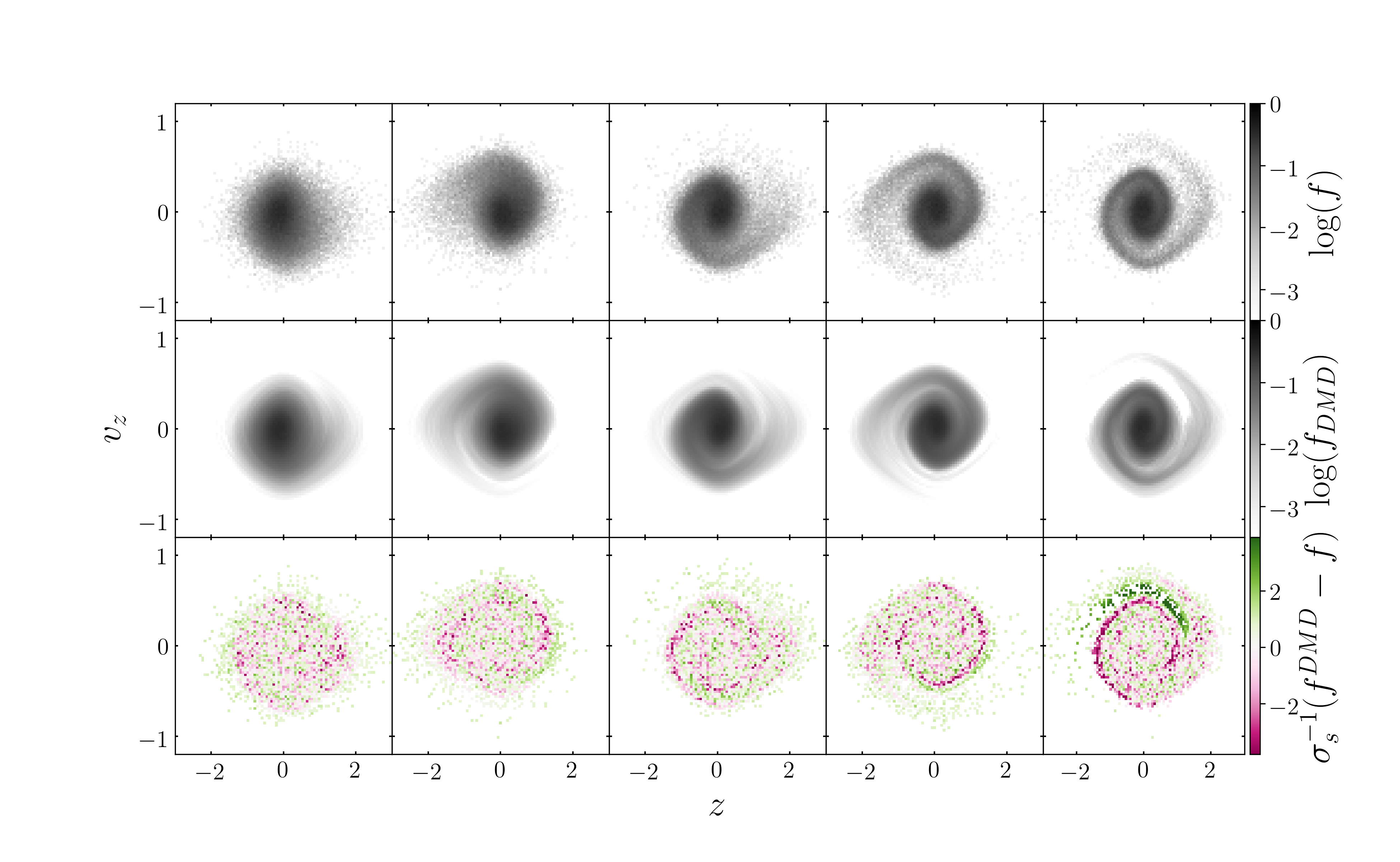}
	\caption{Comparison of the DMD solution and the simulation data with residuals for $ \alpha=0.2 $. The error is computed in the same way as in Fig. \ref{timeevolutionslab}. The snapshots are the same of those in Fig \ref{timeevolution}.} \label{timeevolutionDMD}
\end{figure}

We now apply the DMD algorithm to the isothermal plane simulation data. The data matrices $\mathbf{X}$ and $\mathbf{X'}$ are constructed in the same way as described in Section \ref{slabSimulation}, and the DMD solution is again computed with a rank of $ r=35 $. In Fig. \ref{timeevolutionDMD} we show a comparison of the simulation and DMD solution with residuals for the representative case of $ \alpha=0.2 $. As with the homogeneous slab, the magnitude of the errors are within acceptable values given the noise in the simulation, and there is again some weak systematic structure in the residuals. The systematic structure appears more pronounced here than for the slab, however most of it can be explained by the DMD solution attempting to capture the wispy nature of the simulation along the spiral arm, and essentially over fitting in certain regions. We emphasize that the dominant structure of the system throughout its time evolution is captured well, and for the purpose of extracting the dominant modes, we believe this is sufficient. 

\begin{figure}
	\centering  
	\includegraphics[width=9cm]{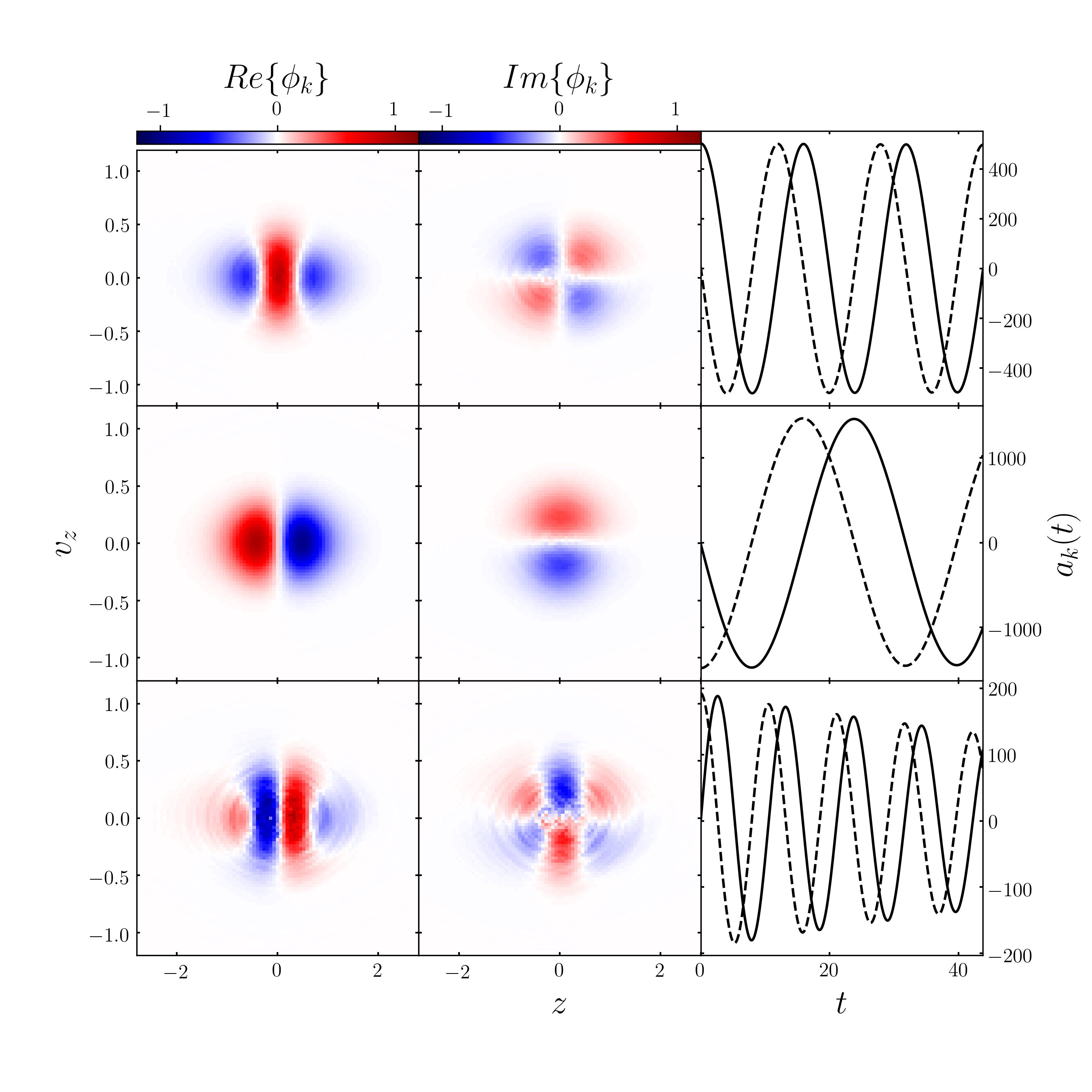}
	\caption{Three most dominant modes aside from the equilibrium mode for the case $\alpha=0.8$. In the right column, the solid and dashed curves correspond to the real and imaginary components of the amplitudes respectively. The modes are evaluated on the same grid used for the histograms in Fig. \ref{timeevolution}. The time axis in the third column is in units of $ 9.8 $ Myr.}\label{livefrac0.8}
\end{figure}

\begin{figure}
	\centering  
	\includegraphics[width=9cm]{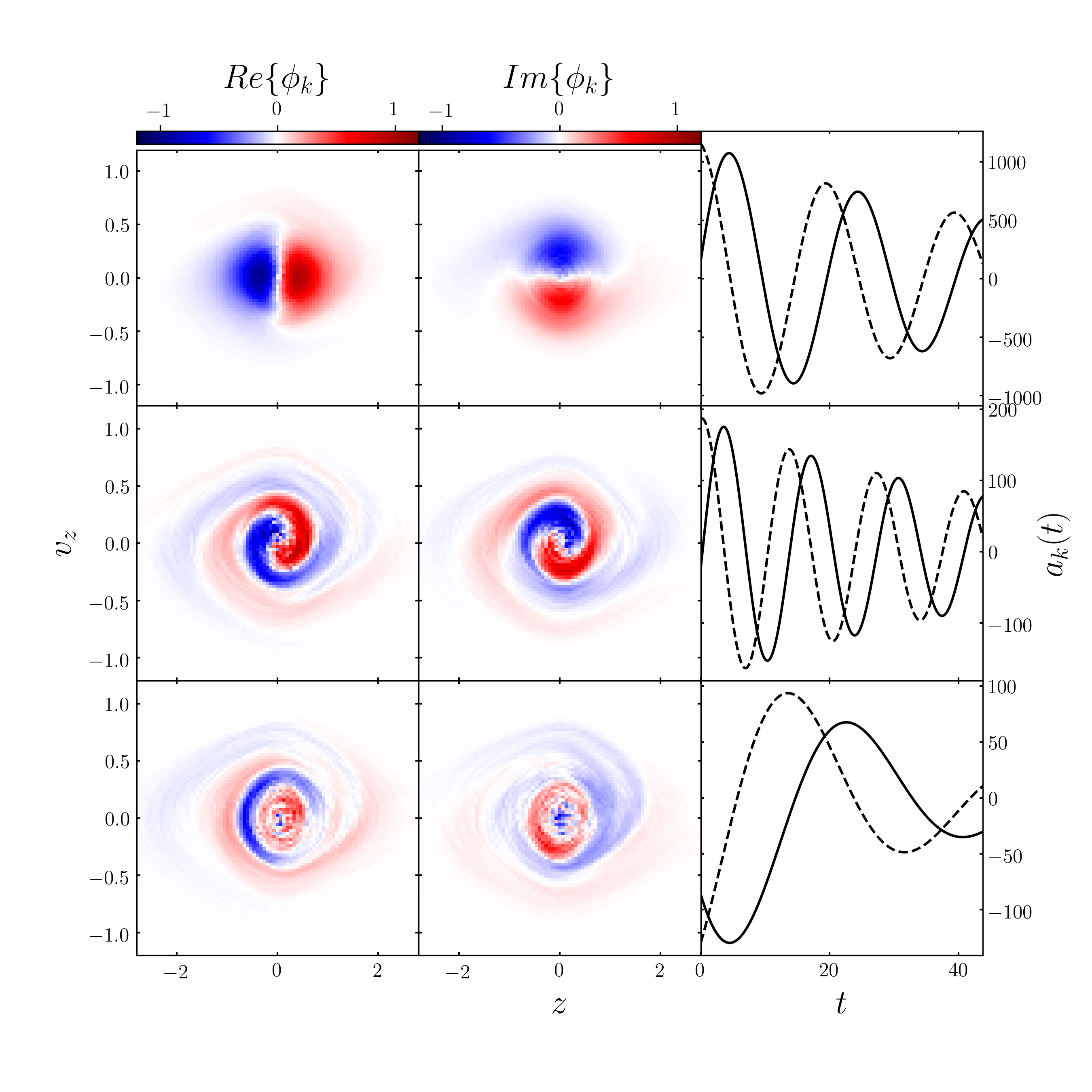}
	\caption{Same as figure \ref{livefrac0.8} for $ \alpha = 0.5 $}\label{livefrac0.5}
\end{figure}

\begin{figure}
	\centering  
	\includegraphics[width=9cm]{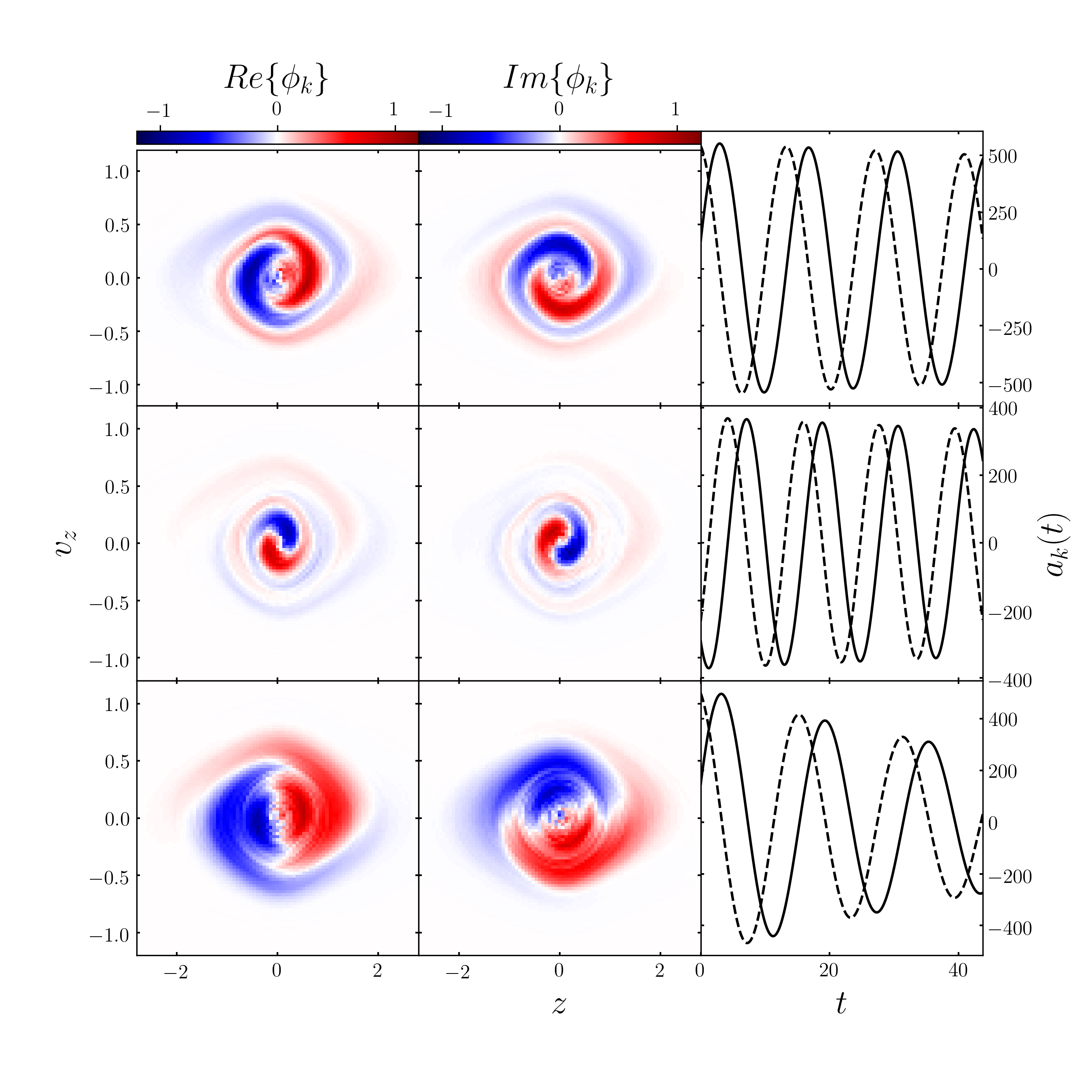}
	\caption{Same as figure \ref{livefrac0.8} for $ \alpha = 0.2 $}\label{livefrac0.2}
\end{figure}

The decompositions for each live fraction yield a zero frequency equilibrium mode, and several complex oscillatory modes. In what follows we focus only on the complex modes. In Figs. \ref{livefrac0.8}, \ref{livefrac0.5}, and \ref{livefrac0.2} we show the three most dominant modes of the DF excluding the equilibrium mode, for the live fractions $ \alpha=0.8,0.5,0.2 $. As these are all complex modes, we show only one member of each conjugate pair. We include the real and imaginary components, as well as the time dependent complex amplitudes. As indicated by the phase difference in the amplitudes, the modes effectively oscillate between the real and imaginary components. 

In general we obtain a combination of damped and un-damped modes. Un-damped and very weakly damped modes are dominant in the DMD solution. These dominant modes behave similarly to normal modes, and in some cases \textit{are} the true modes (like in section \ref{SLAB}). They persist on long time scales relative the dynamical time of the system. Strongly damped modes correspond to transient responses  of the system and the subsequent decay due to phase mixing and Landau damping \citep{binney2008}. 

In addition to the mode structures, we can also make use of the corresponding eigenvalues in understanding the temporal behavior of the modes, and the dimensionality of the DMD solution. Recalling that the frequencies are related to the eigenvalues via $\omega_j = \ln{\lambda_j}/\Delta t$, and considering the eigenvalues in polar form, $ \lambda=|\lambda|e^{i\theta_\lambda} $ one has that the frequencies can be written as  

\begin{equation}\label{frequency}
\omega = \tfrac{1}{\Delta t}\big(\ln|\lambda|+i\theta_\lambda\big).
\end{equation} 

\begin{figure}
	\centering  
	\includegraphics[width=6cm]{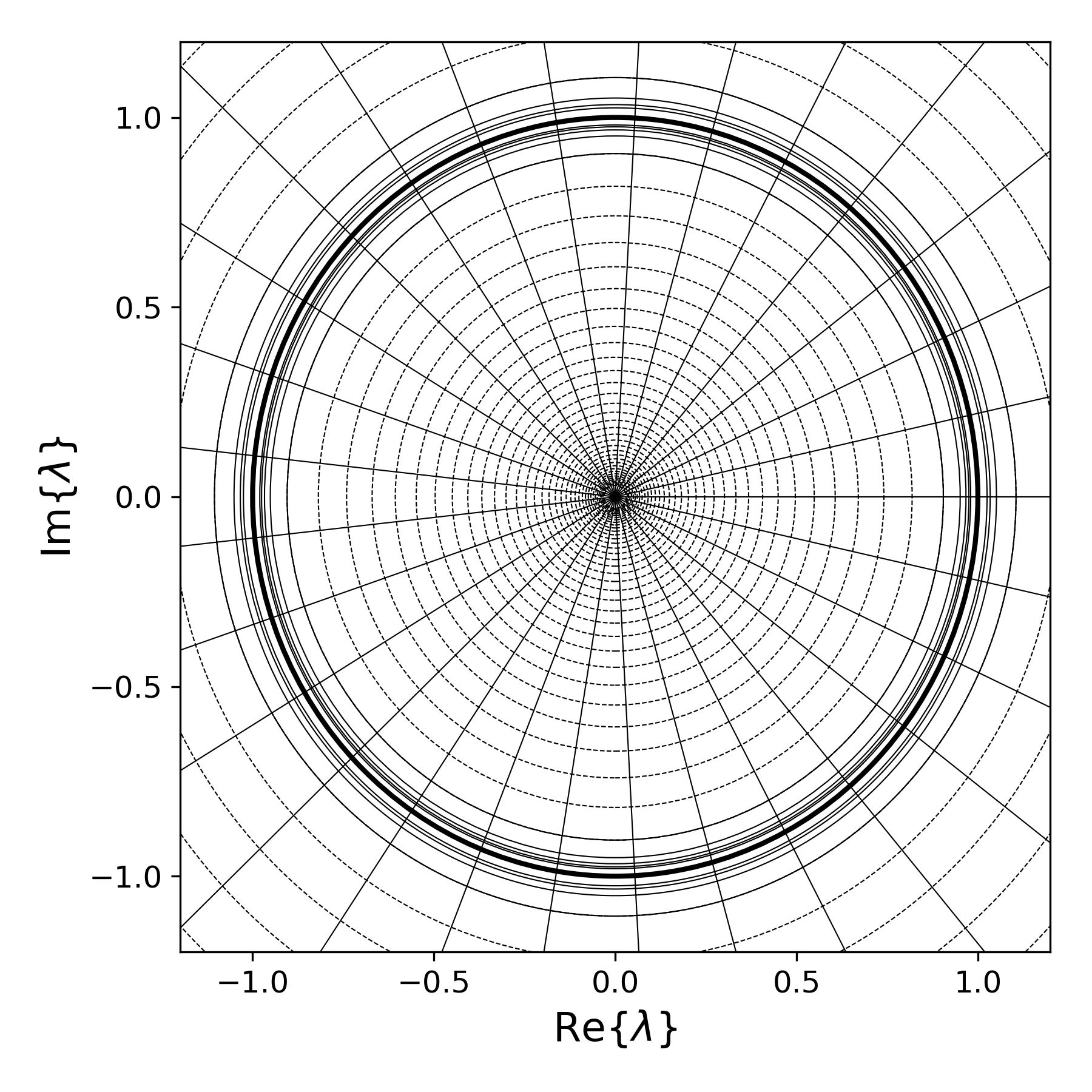}
	\caption{The coordinate system used to interpret the eigenvalues corresponding to the DMD modes. The radial lines are curves of constant mode frequency, increasing with absolute angle in the  plane. The thick circle is the unit circle, and modes with eigenvalues lying on this curve do not grow or decay (true modes). Curves inside the unit circle are curves of constant lifetime for stable modes, with lifetime increasing as they move towards the unit circle. Curves outside the unit circle are curves of constant growth rate for unstable modes, with growth rate increasing as they move away from the unit circle. The solid curves are integer multiples of orbital period, while the dashed curves go as $ \tau=T/s, \ s\in\mathbb{Z} $.}\label{eigenvaluecoords}
\end{figure}

\begin{figure}
	\centering  
	\includegraphics[width=9cm]{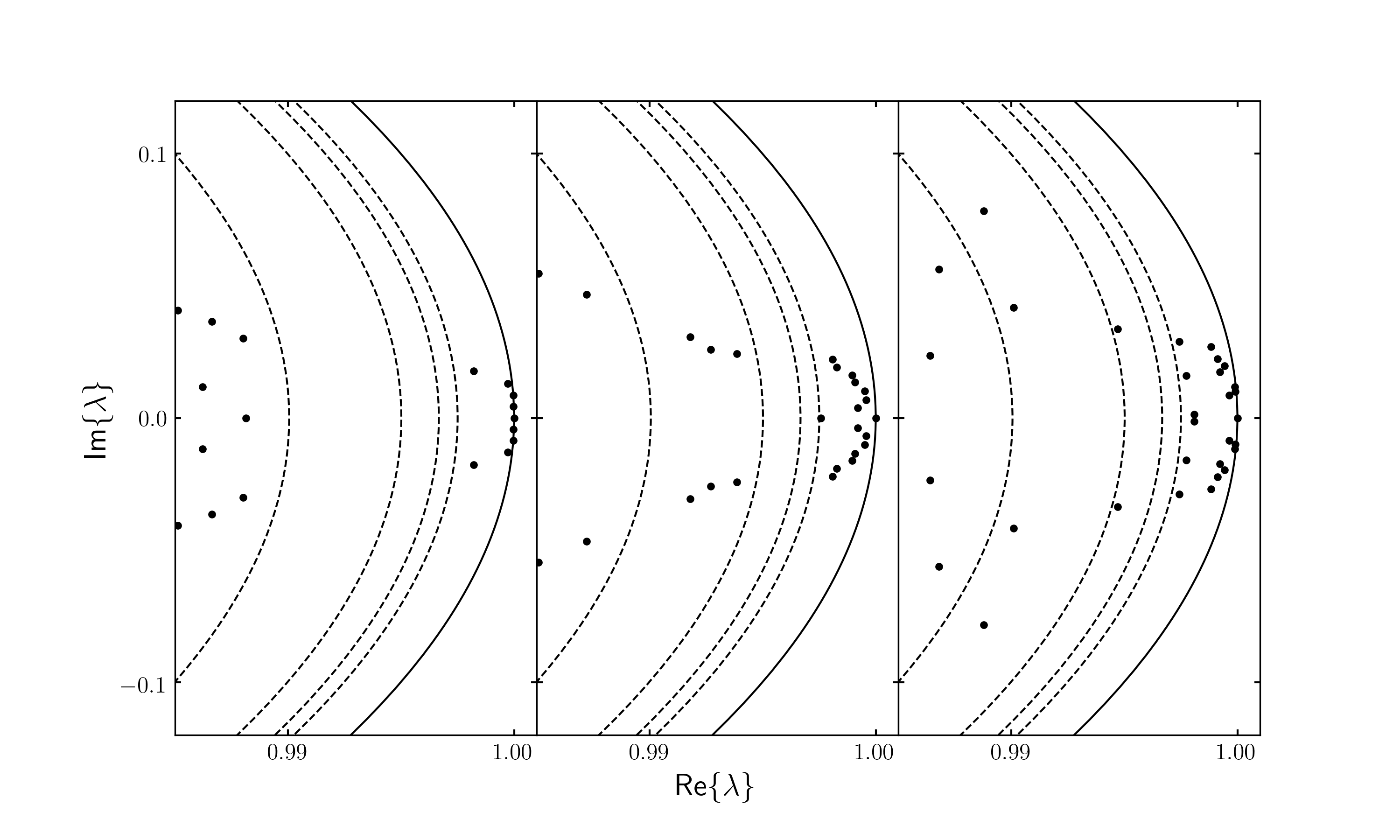}
	\caption{Eigenvalues plotted in the complex plane as described in Fig. \ref{eigenvaluecoords}. From left to right, the panels correspond to $ \alpha=0.8,0.5,0.2 $. The dashed curves here correspond to modes lifetimes of $ T $, $ 2T $, $ 3T $ and $ 4T $ in that order as they approach the unit circle. There are six eigenvalues that are not shown in the $ \alpha=0.8 $, as they are well within the $ \tau = T $ circle, and correspond to extremely rapidly damped modes that do not contribute to the dominant structure.  }\label{eigenvalues}
\end{figure}

\noindent From equation \ref{fseries}, we see that $ |\lambda| $ determines the growth or decay rate of each mode, while $ \theta_\lambda $ sets the mode oscillation frequency. It is also convenient  to define the mode lifetime

\begin{equation}\label{lifetime}
\tau = \frac{\Delta t}{\ln|\lambda|},
\end{equation}

\noindent which we can compare to the orbital period  of the system, $T$. We show in Fig. \ref{eigenvaluecoords} the coordinate system in the complex plane that we will use to interpret the eigenvalues. This consists of curves of constant lifetime and frequency, indicating the physical meaning of an eigenvalue's position in the complex plane. Modes with $ |\lambda|\approx 1 $ will persist, while modes with $ |\lambda|<1 $ will damp on the timescales indicated by the lifetime curves. Thus, the long term behavior of the system may be approximated by a sum over modes with eigenvalues near the unit circle. In Fig. \ref{eigenvalues} we show the eigenvalue spectra for each choice of $\alpha$ on the relevant patch of the plane in Fig. \ref{eigenvaluecoords}. In addition to mapping the modes to lifetime and frequency, these plots provide an indication of the dimensionality of the phase space structure in the DMD basis, as they show how many modes will persist through the time evolution of the system.

Note that in this discussion, we are concerned only with the dynamics present in the simulation. That is, we are using DMD as a diagnostic tool, not for future state prediction (see \cite{kutz2016}). We do not then concern ourselves with the accuracy of the DMD solution outside of the time interval of the simulation. 

\subsection{Mode Interpretation}

We now discuss the physical interpretation of the isothermal plane DMD modes, describing their possible connection to known normal modes in Section \ref{bending}, and their relevance to phase mixing in Section \ref{spiral}.

\subsubsection{Bending and Breathing Modes}\label{bending}

In the case of $ \alpha=0.8 $, where self-gravity dominates the evolution over the external potential, the dominant modes in rows one and two of Fig. \ref{livefrac0.8} share aspects of their morphology with breathing and bending modes derived in \cite{mathur1990} and \cite{weinberg1991} and further studied in \cite{chequers2014} and \cite{widrow2015}. These modes are not damped, and have integer multiple frequencies. The sum of the conjugate pairs for each of these modes rotate clockwise in the $ (z,v_z) $ plane with a pattern speed set by the mode frequency, which is also consistent with theoretical models. However the modes in the first and second rows of Fig. \ref{livefrac0.8} oscillate between the structures shown in the real and imaginary components, which differs from the Weinberg modes. The true modes of the isothermal plane are discrete, but lie very close to a continuum. As such they have proven difficult to excite in simulations \citep{weinberg1991}. We believe that the modes we see here are mixtures of the true modes.  

\subsubsection{Phase Space Spirals}\label{spiral}

In the cases with moderate or weak self-gravity ($ \alpha=0.5 $ and $ \alpha=0.2$, respectively) phase mixing becomes important and the DMD algorithm finds modes with spiral structure. As these spiral modes capture the structure in the DF we expect from phase mixing, one might think they are strictly a transient response. Although this is the case for many of the modes, there are un-damped persisting spiral modes, such as those in rows one and two of Fig. \ref{livefrac0.2}. Additionally, the mode in row one of Fig. \ref{livefrac0.5}, which is qualitatively similar to a bending mode, has a left-handed spiral structure apparent in both the real and imaginary components. Although we do not claim that there are true spiral modes, it is evident that the inclusion of self-gravity in a phase mixing system allows for persistent, dominant spiral modes. Consequently this means phase space spirals can persist on longer timescales than expected from pure kinematic phase mixing, as was argued in \cite{darling2018}. 

The variation in mode structure and lifetime with live fraction indicates that self gravity and phase mixing can be thought of as competing effects, and there is a regime of relative strength of these effects in which spiral modes can persist on long timescales. We argue then that if the live fraction is in the appropriate regime, phase space spirals could be observed at a somewhat arbitrary time in the evolution of the system, independent of the timing expected from pure kinematic phase mixing. Under the assumption that stars in the solar neighborhood evolve according to their mutual interactions in the presence of an external potential, an estimated effective live fraction for this region of the Galaxy could yield an alternative estimate on the timing of GDR2 spirals than those made assuming only a background potential. 

The result that a system undergoing phase mixing in the presence of self gravity can possess persisting spiral modes, in combination with the connection between perturbation theory and DMD demonstrated in Section \ref{SLAB}, indicates that one may consider spiral modes as first order perturbations on the equilibrium distribution. With this, it would be possible to estimate the potential corresponding to perturbative spiral modes, which may be indicative of the form of perturbations that lead to phase space spirals like those observed in GDR2.

\section{Extensions}\label{extension}

Thus far we have focused on one-dimensional models. A logical next step is to apply DMD to full three dimensional simulations. The DMD algorithm itself is robust to very large data matrices. The challenge is to find an appropriate set of observables for each snapshot. In the one-dimensional case, we had the luxury of high particle resolution in the $ (z,v_z) $ phase space, and consequently could use a high grid resolution when evaluating $ f(z,v_z) $. In the case of full three-dimensional simulations, a simple binning procedure in 6D phase space is unfeasible for anything beyond the coarsest grid. 

Suppose one is interested in using DMD to study the phase spirals found by \cite{antoja2018} across the disc. One might imagine the following strategy: First sort particles in $N_R$ radial bins. For each bin, construct a Fourier series in Galactic azimuth, keeping only $N_\phi$ terms. Then, for each azimuthal mode number $m$ bin particles in an $N_z\times N_v$ grid across the $z-v_z$ plane. Finally, for each of the $N_R N_\phi N_z N_v$ cells, compute the first $m_{\rm max}$ moments of $v_R$ and $v_\phi$. For example, with $m_{\rm max}=2$ there are six moments. For $N_R=20$, $N_\phi=9$, $N_z=N_v=40$, and $m_{\rm max}=2$ we have approximately $1.7M$ cells, which would require a simulation with several hundred million particles, a large but feasible quantity.

On the other hand, if one is interested in spiral structure and warps, then an approach akin to the Fourier methods introduced by \cite{sellwood1986} and extended to bending waves by \cite{chequers2018} provide an attractive alternative. In this case, one computes the surface density and vertical moments of the DF, such as the mean midplane displacement and mean vertical velocity as functions of $R$ and $\phi$ across the disc. These quantities in a time series are then used to compose the DMD data matrices.

\section{Conclusion}\label{Conclusion}

We have showed that DMD facilitates an analysis analogous to normal modes, with great generality in how it may be applied to problems in galactic dynamics. When applied to time series measurements of phase space density, DMD yields a finite series solution for the DF. The dominant terms in this series are typically un-damped or very weakly damped oscillations that can persist on long time scales relative to the orbital period. Moreover, the method can capture the physics of both phase mixing and self-gravity. By computing DMD modes, one can describe and study time-dependent phase space structure throughout its evolution in terms of a just a few eigenfunctions and their time-dependent coefficients. This provides a much richer look at the evolution of structure than typical spectral methods and allows analysis in terms of very few quantities compared to the full data set yielded by N-body simulations. The method should be even more powerful in the case of simulations of the complete 6D phase space.

We have observed how the competing effects of self gravity and phase mixing manifest in DMD modes. In the presence of both effects, persisting spiral modes arise. In the DMD solution, the spiral modes are responsible for the apparent phase mixing in the full DF. The eigenvalues associated with these modes should yield insight into the timescale of phase space spirals, and the structure of the eigenfunctions themselves can inform perturbations in the potential associated with non-equilibrium phenomena.

The observational evidence for a Galaxy in disequilibrium has led to a keen interest in the DF for the stellar disc and specifically the form and timescale of the perturbations. Since this is inherently a time-dependent problem, numerical studies are a key component in understanding the complete picture. DMD has the potential to provide valuable insight into stellar dynamics just as it has in the field of fluid dynamics.

\section*{Acknowledgments} {It is a pleasure to thank Martin Weinberg for useful discussions. We also thank the referee, whose comments motivated the inclusion of the homogeneous slab model. LMW acknowledges the hospitality of the Kavli Institute for Theoretical Physics, which is supported in part by the National Science Foundation under Grant No. NSF PHY-1748958. This work was also supported by a Discovery Grant with the Natural Sciences and Engineering Research Council of Canada.}


\bibliographystyle{mnras}
\bibliography{bibliography.bib} 


\bsp    
\label{lastpage}

\end{document}